# The role of VUV radiation in the inactivation of bacteria with an atmospheric pressure plasma jet

Simon Schneider, Jan-Wilm Lackmann, Dirk Ellerweg, Benjamin Denis, Franz Narberhaus, Julia E. Bandow, and Jan Benedikt*

__________

S. Schneider, D. Ellerweg, Jun.-Prof. Dr. J. Benedikt
Coupled Plasma-Solid State Systems, Department for Physics und Astronomy, Ruhr-University Bochum, Universitätsstraße 150, 44780 Bochum, Germany
E-mail: jan.benedikt@rub.de

J.-W. Lackmann, Prof. Dr. F. Narberhaus, Jun.-Prof. Dr. J. E. Bandow
Microbial Biology, Department for Biology und Biotechnology, Ruhr-University Bochum, Universitätsstraße 150, 44780 Bochum, Germany

B. Denis
Institute for Electrical Engineering and Plasma Technology, Ruhr-University Bochum, Universitätsstraße 150, 44780 Bochum, Germany

__________

A modified version of a micro scale atmospheric pressure plasma jet (µ-APPJ) source, so-called X-Jet, is used to study the role of plasma generated VUV photons in the inactivation of *E. coli* bacteria. The plasma is operated in He gas or a He/$O_2$ mixture and the X-Jet modification of the jet geometry allows effective separation of heavy reactive particles (such as O atoms or ozone molecules) from the plasma-generated photons. The measurements of the evolution of zone of inhibitions formed in monolayers of vegetative *E. coli* bacteria, of VUV emission intensity and of positive ion spectra show that photochemistry in the gas phase followed by photochemistry products impacting on bacteria can result in bacterial inactivation. Interestingly, this process is more effective than direct inactivation by VUV radiation damage. Mainly protonated water cluster ions are detected by mass spectrometry indicating that water impurity has to be carefully considered. The measurements indicate that the combination of the presence of water cluster ions and $O_2$ molecules at the surface leads to the strongest effect. Additionally, it seem that the interaction of VUV photons with effluent of He/$O_2$ plasma leads to enhanced formation of $O_3$, which is not the case when only $O_2$ molecules and gas impurities at room temperature interacts with plasma generated VUV photons.

# Introduction

Cold atmospheric pressure plasmas (CAP) are increasingly in the focus of researchers investigating their possible applications in medicine or the food packaging industry.[1, 2] CAP jets are able to inactivate bacteria or fungi or damage bio-macromolecules.[3, 4, 5] They offer therefore an alternative to standard sterilization methods there, where thermo-labile and vacuum-sensitive objects (plastics or living tissues) have to be treated. In addition, several studies investigate the influence of CAP treatment on wound healing and cancer cells.[6, 7] These plasmas produce positive and negative ions, (V)UV radiation, and reactive radical species, which interact with the treated surface. The effects of different plasma-generated species on the treated systems are a topic of current scientific discussions.[1] The role of reactive oxygen species (ROS) has been stressed by several authors as key molecules affecting vegetative prokaryotic and eukaryotic cells.[8, 9] More recently, the combination of ions and ROS has also been discussed.[2, 10] Capacitively coupled atmospheric pressure plasma jet (APPJ) sources operated with He with some addition of $O_2$ ($\leq 1\%$) are known to be efficient sources of ROS, particularly oxygen atoms, ozone molecules ($O_3$), or singlet delta oxygen metastables $O_2(a^1\Delta_g)$. Measurements and modeling have been reported for a coaxial jet with a 1 cm diameter inner electrode and a 1 mm electrode gap, a parallel plate jet with 1 mm electrode separation and 1 mm electrode width, or for sources with an electrode width larger than 1 mm.[11, 12, 13] The plasma dynamics and plasma chemistry in these discharges have also been modeled by several authors.[14, 15] These works show that densities of above mentioned ROS are around $10^{15}$ cm$^{-3}$ in the effluent of these jets and can be tuned by adjusting $O_2$ concentration, applied power, gas flow, and jet-substrate distance. This kind of plasma is a promising tool for treatment of living tissues or for antibacterial treatment of surfaces at atmospheric pressure. The knowledge of ROS densities and, therefore, also the fluxes could be used to evaluate quantitatively the effects of ROS for example on vegetative bacteria. With respect to the surface being treated, these sources are remote sources. The plasma is confined between electrodes and only a neutral plasma effluent without charged species reaches the surface. It is probably the absence of the flux of charged species to the surface that makes the treatment of bacteria or living tissues less effective than in low-pressure reactors where the substrate is in direct contact with the plasma. However, it allows the fundamental study of the effects of different plasma components on the living cells. An unknown factor is the effect of VUV photons, which are possibly produced in the plasma in addition to reactive particles. These photons can propagate unabsorbed through He atmosphere, irradiate the treated surface and induce uncontrolled radiation damage. We report in this article the results of a study on the role of VUV photons in inactivation of vegetative *Escherichia coli* bacteria. This study is performed with a modified microplasma jet, which allows well-defined separation of plasma-generated VUV photons and reactive particles.

**Experimental Setup**

The µ-APPJ has a very simple geometry. It is formed by two stainless steel electrodes (length 30 mm, thickness 1 mm) with a separation of 1 mm, and two glass plates, which are glued to the electrodes on the side and confine the inter electrode volume. The well-controlled flow conditions without admixture of surrounding ambient atmosphere and a very good optical access to the plasma is maintained in this way. The plasma is generated in He gas flow of up to 5 standard liters per minute (slm) with small concentrations (<1.5 %) of some reactive gas (in this case $O_2$). A typical α-mode discharge is formed when sinusoidal driving voltage with root-mean-square value of 150-270 V (frequency 13.56 MHz) is applied to the electrodes. This source has been described and studied in the past and quantitative measurements of O and $O_3$ densities as function of $O_2$ concentration, applied power, and distance to the jet are available.[11, 16, 17] Additionally, this jet geometry with He/$O_2$ gas mixture has been modeled and discussed in the literature.[14, 15] The gas temperature in the plasma effluent has also been measured for a 1% $O_2$ admixture and it was below 34◦C.[11] The temperature is slightly higher at lower $O_2$ concentrations but drops quickly as the distance from the jet increases.

A modified µ-APPJ, a so-called X-Jet, is used in this work. The nozzle of the µ-APPJ is extended by two crossed channels as shown in a photograph in **Figure 1a**. These channels are formed from glass and metal building blocks with 1 mm thickness, and fixed between two glass plates together with both electrodes. One channel is a direct extension of the inter-electrode region (direct channel) and the other channel (side channel) crosses the direct channel under a 45 degree 3 mm downstream of the end of the electrodes. Both channels have the same 1x1 mm2 cross section. An additional He flow is applied to the side channel to divert particles in the plasma effluent from the direct channel into the side channel. This is possible, because the particle transport is controlled mainly by convection at atmospheric pressure and the flow velocities used. The diffusion is less important due to high collision rates. Contrary to particles, VUV and UV photons generated in the plasma can propagate further through the direct channel (also filled with He), cf. the scheme in **Figure 1b**. Only a very small fraction (<1.5%) of photons is reflected or scattered into the side channel. Therefore, operation of the X-Jet with additional He flow through the side channel leads to effective separation of the plasma-generated reactive particles (for example O atoms or $O_3$ molecules which are emanating from the side channel) and plasma generated photons, which propagate through the direct channel. Additionally, the experiments are performed in a chamber with controlled He atmosphere (volume ~ 8 liters) to minimize the influence of ambient atmosphere. The substrate was always placed perpendicular to the axis of the corresponding channel used for treatment at a distance of 4 mm. The details regarding the separation of reactive particles and photons, the results of the simulation of the convection/diffusion transport in the X-Jet, and the testing of the X-Jet performance in etching experiments of plasma polymer films, emission spectroscopy measurements in the 115 - 875 nm wavelength range, as well as treatment of bacteria in their vegetative form can be found elsewhere.[18]

Combined and separate effects of the reactive heavy particles and the VUV and UV radiation of the plasma effluent on bacteria or other substrates can now be studied in the following ways: i) photons and reactive particles are applied together: An X-Jet without additional He flow in the side channel will result in the transport of both reactive particles and photons through the direct channel. ii) Reactive particles only: The same He flow is used in both channels. The additional flow through the side channel will push the heavy particles from the plasma effluent into the side channel as demonstrated in Figure 1b. The flow rates through both channels after the crossing are the same due to the symmetry of this geometry. The flux of ROS at the exit of the side channel is expected to be similar to the ROS flux at the direct channel in i), although some differences will occur due to a missing photo-dissociation and excitation of ROS and $O_2$ after the crossing of both channels (see also discussion of *E. coli* treatment later) and due to a slightly asymmetric velocity field across the side channel. The plasma generated VUV and UV photons cannot enter directly into side channel due to geometry constrains, which was corroborated by measurements presented later in this article. iii) VUV and UV only: With the additional He flow, only VUV and UV photons but no ROS exit the direct channel. Higher He flow in the side channel can be used to make sure that no ROS from the plasma diffuse into the direct channel.

**Measurement of emission intensity**

A solar blind VUV and UV detector (PMT-142, effective in the 115 - 450 nm wavelength range with a maximum relative efficiency at around 220 nm) in an evacuated housing with $MgF_2$ window has been used to measure the wavelength integrated intensity of VUV and UV emission from the direct channel of the X-Jet.[19] A 1 mm diameter diaphragm was placed on the $MgF_2$ window and the jet was always at 4 mm distance from the window to maintain the same acceptance angle for each measurement. The same setup has been used in previous measurements to verify that the photon flux (in this wavelength range) through the side channel is negligible.[18]

**Measurement of ion fluxes**

Molecular beam mass spectrometry (MS) sampling system is used to measure positive ions in the gas phase. The three stages differentially pumped MS sampling system described previously is used for these measurements.[17, 20] This system is equipped with a rotating chopper to modulate the molecular beam and maintain vacuum in the MS. This chopper was, however, not used here because a smaller sampling orifice (20 μm diameter) was installed. The X-Jet is placed into a small chamber with controlled He atmosphere, which is mounted directly at the mass spectrometer front plate with the sampling orifice. Comparable conditions as during the treatment of bacteria are arranged in this way.

**Preparation of biological probes**

Vegetative *E. coli* cells have been used as model substrate in this study. The following procedure has been used to prepare the substrate and evaluate the effects of exposure to plasma effluents. *E. coli* K12 liquid cultures were incubated for 18 hours over night at 37 °C in LB medium.[21] The cultures were diluted to an optical density of 0.05 at 580 nm and were sprayed onto LB agar plates for 1 second. A monolayer surface coverage with $\sim 2.2\times10^3 \cdot cm^{-2}$ cell density is achieved in this case. The plates were grown for 1 hour at 37°C before plasma treatment. After plasma treatment, the sample plates were incubated over night for 18 hours at 37°C to allow survivors to grow. Zones of inhibition were observed where treatment was lethal. It was checked that treating of the agar plates with plasma before the application of cells had no effect on bacterial growth. Furthermore, no pH change of the medium occurred during plasma treatment. We report in the following results of treatment of vegetative *E. coli* bacteria by different components of plasma effluent. The main focus is on the role of VUV and UV photons in the inactivation of bacteria.

**Experimental Results and discussion**

The big advantage of the X-Jet is that it allows us to separate the effects of plasma-generated VUV and UV photons from the effects induced by reactive particles ($O_3$, O, impurities,...). The effects of VUV and UV photons only, reactive particles only, and photons and reactive particles together have already been studied in *E. coli* and selected results are shown in **Figure 2**.[18] The following observations have been made: i) the VUV and UV photons generated in plasma had only a weak effect on *E. coli* survival. No inactivation was visible in samples treated with the direct channel for 1 and 3 min of the treatment. The zone of inhibition at the jet axis appeared only after 6 min of the treatment with photons. ii) The combined treatment and reactive particle-only treatment showed typical dose-effect relationships. Elongated treatment times resulted in larger zones of inhibition and lower numbers of colony forming units. In both cases, the whole area of the Petri dish (80 mm diameter) was affected after 3 min of treatment. iii) Inactivation by the combined treatment was approximately twice as fast as with the reactive particle-only treatment (cf. Figure 2). The fast inactivation by reactive particle-only and by the combined treatment is due to the fact that the μ-APPJ is an effective source of atomic oxygen and ozone. For example, the densities (concentrations) of O and $O_3$ measured by molecular beam mass spectrometry at 4 mm distance from the jet under conditions used in this work are $7\times10^{14}$ cm$^{-3}$ (~28 ppm) and $5\times10^{14}$ cm$^{-3}$ (~ 20 ppm), respectively.[17] The concentration of ozone increases with the distance and reaches ~56 ppm at 50 mm. The concentrations of both O and $O_3$ can be expected to be high enough to cause the observed effects. Ozone is known for its bactericidal activity. Just 5 min of treatment with 0.2 ppm of ozone in water is lethal for *E. coli*, *Bacillus cereus*, or *Bacillus megaterium*.[22] Moreover, experiments in air with ozone have shown that it also effectively kills bacteria on agar plates. Ozone concentrations below 1 ppm and treatment times smaller 100 min have been reported to be effective in killing *Staphylococcus albus*, *Streptococcus salivarius*, and *Bacillus prodigiosus*.[23] Atomic oxygen also has detrimental effects on bacteria. It can etch biological material or cause oxidative stress inside the cell. We have shown that the flux of atomic oxygen from the side channel of the X-Jet etches a BSA protein layer or a model plasma polymer film of hydrogenated amorphous carbon with a rate of 30 nm/min and that the area on the surface, which is affected by atomic oxygen, is limited to a diameter of ~10 mm.[18, 24] This limitation is given by the fast loss of O in the gas phase in the three body reaction:[25]

$$O + O_2 + He \rightarrow O_3 + He \tag{1}$$

Regarding the fact that in treatments with reactive particles longer than 6 min the whole Petri dish (80 mm diameter) was affected and taking into account the short lifetime of O atoms, we conclude that mainly $O_3$ is responsible for the inactivation of bacteria at large distances from the jet axis. Simultaneous action of O, $O_3$, $O_2$ metastables, some possible impurities (like OH from water) and, in the case of combined treatment through the direct channel, the VUV and UV radiation most likely play a role in bacterial inactivation near to the jet axis.[18]

An interesting effect, which is further studied and discussed in this article, is the fact that the combined treatment shows effects twice as fast as the treatment with reactive particles without photons. It was expected to see more effective inactivation when reactive particles and photons were treating the substrate simultaneously. However, the simultaneous treatment at the surface is limited only to an area of 2 to 3 mm diameter just underneath the nozzle of the direct channel. The rest of the substrate is shadowed by the channel structure and no synergistic effects were expected there. However, Figure 2 shows that the zone of inhibition of the combined treatment had a diameter of 20 mm after 1 min of treatment whereas the treatment with reactive particles only, resulted in a 5 mm diameter after the same time. Additionally, the combined treatment at 3 min has an effect comparable to the reactive particles-only treatment at 6 min. These observations indicate that the plasma effluent is changed by the presence of VUV photons. Some photochemistry reactions take place in the gas phase in the second part of the direct channel and on the way to the substrate, which are missing in the reactive particle-only treatment. Additionally, the fact that this change takes place at greater distance from the jet axis hints that additional ozone might be formed. The X-Jet geometry offers a unique opportunity to study these effects in more detail.

**Study of the photochemistry in the X-Jet**

The interaction of photons with some particles in the gas phase and its effects on bacteria can be studied by admixture of these particles into the He flow administered through the side channel. Here, we tested this using $O_2$ as it has the second highest density in the plasma effluent (0.6% concentration, consumption below 5% at $U_{RMS}$ = 230 V) after He and its photochemistry leading to the formation of ozone in the stratosphere is well known. **Figure 3** shows the comparison of zones of inhibition induced by photons only (He gas is supplied through the side channel) and by photons and $O_2$ photochemistry products (He gas with admixture of $O_2$ at different concentrations is fed through the side channel). Conditions in the plasma are the same (He gas flow 1.4 slm, $O_2$ concentration of 0.6 %, $U_{RMS}$ = 230 V). The gas flow in the side channel at s slm is higher than that of the flow exiting the plasma to make sure that diffusion of heavy reactive species from the plasma into the direct channel after the channel crossing is negligible. Treatment times of 1 and 4 min have been used. No inhibition zones are observed after 1 min of the treatment when no $O_2$ is added into the He flow through the side channel. Zones of inhibition with a 3.8 mm diameter appear after 4 min of treatment. These results are consistent with observations presented in Figure 2. Adding 0.8, 4.2 or 20% of $O_2$ into the He gas flow in the side channel always results in the formation of an inhibition zone with diameters between 3 and 4 mm. No clear dependence on the amount of added $O_2$ is observed after 1 min of the treatment. At 4 min diameters increase to 5, 6.5, and 8.5 mm in an $O_2$ concentration-dependent manor. The same trend was also observed in several additional measurements with different plasma conditions and gas flows. The results shown in Figure 3 provide direct evidence that photochemistry of $O_2$ in the effluent of the µ-APPJ has a measurable effect on the inactivation of bacteria on the time scales of minutes.

Based on these result the possible photochemical processes can be discussed. The photodissociation threshold of $O_2$ molecules is at 5.12 eV (wavelength 242.4 nm) and the photoionization threshold at 12.07 eV (102.7 nm).[26] An absolutely calibrated emission spectrum of the jet has been measured in the past down to the wavelength of 115 nm, the cut-off limit of the $MgF_2$ window.[27] Two atomic oxygen emission lines at 115 nm ($^1D$ - $^1D^o$) and 130 nm ($^3P$ - $^3S^o$) dominate the spectrum. Additionally, a weak emission due to parts of the Schumann-Runge bands of $O_2$ and a weak H line at 120 nm ($^2P$ - $^2S^o$) have been observed. The photons with wavelengths of 115 nm and 130 nm can dissociate $O_2$ molecules. The photoabsorption cross-section (which is mainly due to photodissociation) of $O_2$ has its maximum at 140 nm ($2\times10^{-17}$ cm$^2$), its value at 130 nm is only $\sim3\times10^{-19}$ cm$^2$.[26] The mean free path of 130 nm photons is, therefore, more than 20 cm in He atmosphere with only 0.6% of $O_2$. This is corroborated by the measurements of the VUV emission intensity by the solar blind detector sensitive in the 115 – 450 nm wavelength range. The intensity is measured at the direct channel of the X-Jet operated with the additional He flow through the side channel. The change of the emission intensity in the 115 – 450 nm wavelength range as function of the $O_2$ concentration in the additional He flow ($O_2$ concentration in the plasma was kept constant at 0.6 %) is shown in **Figure 4**. Indeed, only ~ 3 % of the light is absorbed on the 7 mm distance from the channel crossing to the $MgF_2$ window if 0.6% of $O_2$ are added into the additional He flow. The absorption increases almost linearly with increasing $O_2$ concentration. Additionally, the radiances of about 10 µWmm$^{-2}$sr$^{-1}$ were measured for both atomic oxygen lines at the distance of 4 mm from the nozzle of the µ-APPJ operated with 0.6 % of $O_2$ in the gas mixture.[27] The radiance increases at lower $O_2$ concentrations in the plasma. It is around 30 µWmm$^{-2}$sr$^{-1}$ for the 115 nm line and around 15 µWmm$^{-2}$sr$^{-1}$ for the 130 nm line at an $O_2$ admixture of 0.1 %.[27] 10 µWmm$^{-2}$sr$^{-1}$ at 130 nm wavelength is approximately a flux of 4×10$^{13}$ cm$^{-2}$s$^{-1}$ photons. This flux is orders of magnitudes smaller than a simulated flux of 8×10$^{16}$ cm$^{-2}$s$^{-1}$ of O-atoms to the surface. Taking into account that only a small fraction of these photons is absorbed (see Figure 4) it is highly improbable that they are responsible for the faster inactivation observed in Figure 2.

Photons below the 115 nm threshold, which could unfortunately not be measured with our diagnostics due to the cutting wavelength of the MgF$_2$ window, can be responsible for this effect. The measurements of other authors on another plasma source operated with He gas show that the He$^*_2$ excimer continuum in the 58-100 nm range or a strong atomic oxygen line at 98 nm can radiate from the plasma.[28] Moreover, the absorption cross section for photoionization of O$_2$ is ~2×10$^{-17}$ cm$^2$ in the 30-100 nm range giving a mean free path of around 3 mm in the gas mixture of He with 0.6 % of O$_2$.[26] The following experiment has been performed to get new insights into the photochemistry of O$_2$ in the effluent. The He gas without addition of O$_2$ was used as a plasma forming gas in the direct channel of the X-Jet. The excimer continuum is expected to be the most intense in this case, because the quenching of the He$^*_2$ in collisions with O$_2$ molecules is reduced. Moreover, no O$_3$ is produced in the plasma. The chamber with controlled He atmosphere is, therefore, not filled with O$_3$, which otherwise emanates from the side channel, and experiments with longer exposure time can be performed. The He flow with variable concentration of O$_2$ (total flow 2 slm) is used again in the side channel of the X-Jet. *E. coli* bacteria are than treated for 1, 4, and 10 minutes in the same way as in Figure 3. The resulting zones of inhibition are shown in **Figure 5**. Very weak effect is seen after 1 min within a 3.2 mm diameter. The effect gets stronger after 4 min, but the affected area is similar. This observation can be explained by direct inactivation of bacteria by VUV photons from the plasma. A relatively large area with 7-8 mm diameter is inactivated after 10 min of treatment, which cannot be explained by the direct effect of photons only. Some reactive species have to be generated in the gas phase (or eventually on the irradiated surface) that are transported with the gas flow further away from the axis of the direct channel. Addition of $O_2$ into the He flow in the side channel has the following effects: i) faster inactivation in a larger area is visible after 1 min and 4 min of the treatment compared to using He gas only, where the higher $O_2$ concentration is more effective, and ii) the affected area is clearly larger (diameter > 5 mm) and inactivation faster when 4.2 % of $O_2$ are added into the He flow. Surprisingly, a zone of inhibition with a diameter of ~7 mm and very sharp boundary is formed after 10 min of treatment in all cases.

These results corroborate again that VUV photons generated by plasma interact with molecules in the gas phase in such a way that resulting products inactivate *E. coli* bacteria. These photochemistry is more effective when the plasma is operated only in He gas without 0.6% of $O_2$ (compare 1 min treatment in Figure 3 and 5). $O_2$ molecules are clearly involved in the inactivation process or in the photochemistry, because the observed effect is larger at higher $O_2$ concentrations in the side channel. However, the fact that prolonged treatment leads to inactivation only in a limited area indicates that O atoms (which would finally react with $O_2$ forming stable ozone) are not a product of this photochemistry. Generation of ozone during the prolonged treatment would lead to inactivation on much larger area, as observed in Figure 2 for the combined treatment and treatment with reactive particles only. This is clearly not the case here. Moreover, the 7 mm diameter zone of inhibition is formed even when no $O_2$ is added into the gas flow. The photoionization of $O_2$ molecules and gas impurities (mainly water molecules) is a plausible explanation of results in Figure 5. Ions can be produced by VUV radiation, they have a short life time due to the recombination with electrons (and can therefore reach only a limited area on the surface, which will be similar for all ions), and the results from the literature indicates that they can act on the bacteria on the treated surface.[10] The molecular beam mass spectrometry can be used to detect these ions. **Figure 6** shows mass spectra of positive ions measured under the conditions from Figure 5. The X-Jet was placed into a small chamber with controlled He atmosphere to simulate the conditions during the treatment of bacteria. The ion spectra are for all conditions dominated by protonated water clusters with 4 (mass 73 amu) and 5 (mass 91 amu) water molecules. No or very weak signals only at the noise level (below 10 count/s) are detected below the mass of 70 amu. It is apparent that water can be ionized effectively by VUV photons generated by He plasma and therefore that we have VUV photons with wavelengths shorter than 115 nm. The water is an impurity in the chamber and also in the gas lines and in the X-Jet (both during the treatment of bacteria and during the MS measurements). It adsorbs at the surface when the chamber is open and no gas flow is applied and desorbs after the start of the experiment. The measurements shown in Figure 6 are done just few minutes after closing the chamber and the start of the gas flows and plasma (which is the situation during the treatment of bacteria). The signal intensities of all ions drop approximately 200 times within one hour of continuous gas flushing and plasma operation. Addition of $O_2$ into the side channel leads to higher relative intensities of water clusters with 5 water molecules (masses 90 and 91) and of ions detected at masses of 95-97 and 114-116. The overall signal is weaker for the conditions with 4.2% of $O_2$ in the side channel, but part of this decrease is probably due to decreasing water concentration in the system over time. A Signal at the mass 32 ($O_2^+$ ion) appears after the addition of $O_2$ molecules into the side channel, but it is very weak at the level below 10 counts/s. The direct photoionization of $O_2$ molecules seems therefore to play only a minor role. The MS measurements can now be related to the results in Figure 5. The zone of inhibition with 7 mm diameter after 10 min of the treatment with He gas without addition of $O_2$ is probably a result of the presence of water, water cluster ions or eventually also water fragments (such as OH). These ions or water fragments can be transported with the gas flow and therefore reach areas which are not irradiated by the VUV photons. The MS measurement does not show significant formation of new positive ions (such as $O_2^+$) by addition of $O_2$. More probably, the $O_2$ molecules reacting with the ions on bacteria can accelerate the effect, therefore explaining the faster inactivation rate and larger inhibition zone diameters after 1 and 4 min of treatment. This observation corroborates the results of Dobrynin *et al.,* who concluded that ions in the presence of oxygen (or reactive oxygen species) have the highest efficiency to inactivate bacteria.[10] It is also in agreement with their conclusion that the presence of non-liquid water leads to the fastest inactivation. In our case, water which is ionized by the VUV radiation can originate directly from the surface of the wet agar medium or vegetative bacteria.

The results presented here are still preliminary. The MS measurements and the treatment of bacteria should be performed under conditions with better control of the water impurity level in the He gas and with the more precise investigation of the time variation of the process. These measurements are out of the scope of this article at the moment. Still, the results presented here provide enough information to demonstrate that VUV photons, photochemistry products and its combination with $O_2$ molecules on bacteria can play an important role in the atmospheric pressure plasma sterilization.

**Conclusion**

A modified version of a micro scale atmospheric pressure plasma jet (µ-APPJ) source, so-called X-Jet, has been used to study the role of VUV photons in the inactivation of *E. coli* bacteria. The plasma was operated in He gas and He/$O_2$ mixture and the X-Jet modification of the jet geometry allows effective separation of heavy reactive particles (such as O atoms or ozone molecules) from the plasma generated radiation. The results clearly show that plasma generated VUV photons play a role in the inactivation of bacteria. However, the photochemistry of the gas phase species (or adsorbed particles) followed by the reaction of photochemistry products with bacteria seems to be the more important effect than direct inactivation of cells by VUV radiation damage. Our results indicate that photoionization of water clusters by plasma generated VUV photons and the subsequent interactions of these ions with bacteria could explain the effects observed. The addition of $O_2$ into the gas accelerates the inactivation, however, only very limited photoionization of $O_2$ is detected. It indicates that a parallel interaction of water cluster ions (and related photochemistry products) and $O_2$ molecules at the surface accelerates inactivation of bacteria. Additionally, the results also indicate that interaction of VUV photons with the effluent of He/$O_2$ plasma (containing for example vibrationally excited molecules or $O_2$ metastables) leads probably to enhanced formation of $O_3$, which is not the case when only $O_2$ molecules and gas impurities at room temperature interact with plasma generated VUV photons.

Acknowledgements: The authors thank Volker Schultz-von der Gathen for fruitful discussions about the operation of the µ-APPJ source. This work has been performed with the support of the research group FOR1123 approved by the German Research Foundation (DFG). This work has also been supported by the Research Department Plasmas with Complex Interactions of the Ruhr-Universität Bochum as well as through a stipend to J.-W. L. from the Ruhr University Research School.

Received: ((will be filled in by the editorial staff)); Revised: ((will be filled in by the editorial staff)); Published online: ((please add journal code and manuscript number, e.g., DOI: 10.1002/ppap.201100001))

Keywords: atmospheric pressure plasma jet; bacterial inactivation; mass spectrometry; photochemistry; plasma sterilization

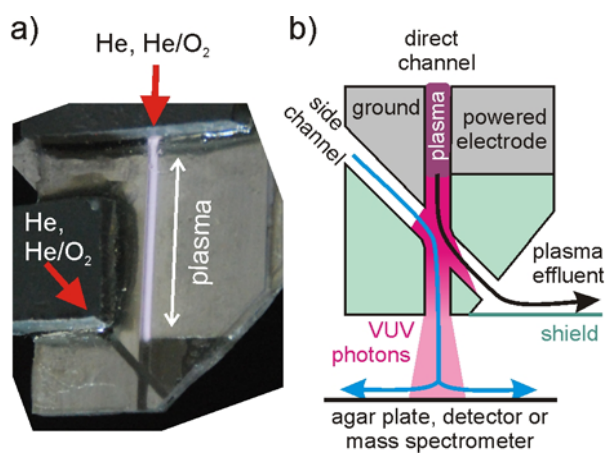

Figure 1. a) Photograph of the µ-APPJ source with two crossed channels after the nozzle, the X-Jet modification. Additional He (or He/O$_2$) flow diverts the plasma effluent into the side channel. VUV and UV photons propagate in line-of-sight with the plasma through the direct channel. b) Schematic representation of the gas flows and photon flux in the channel structure. The overlap of the blue line with the pink region is the area, where the photochemistry takes place.

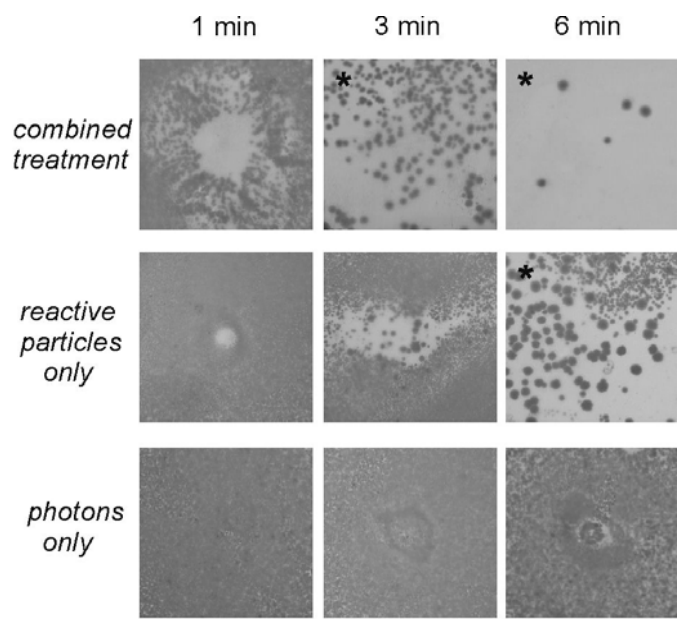

Figure 2. 40 by 40 mm details of photographs of Petri dishes with zones of inhibition in *E. coli* monolayers after 1, 3, and 6 min of treatment with reactive particles only, VUV and UV photons only, or treatment with both combined. Details marked by *: the entire plate (80mm diameter) was affected. Adopted from [18].

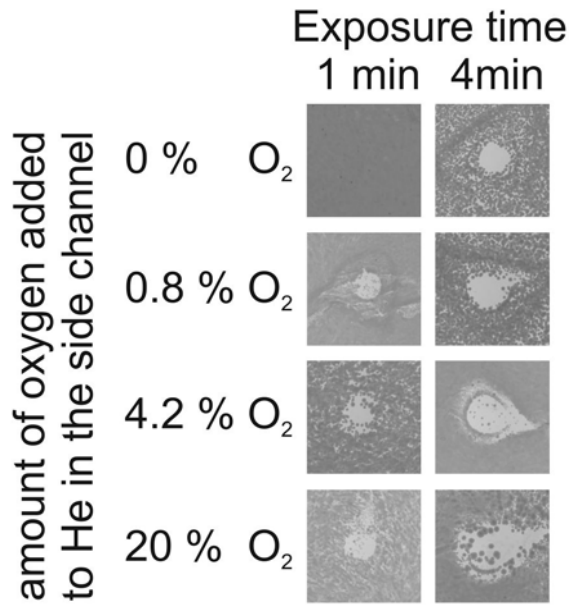

Figure 3: 15 by 15 mm details of photographs of Petri dishes with zones of inhibition in *E. coli* monolayers after 1 and 4 min of treatment. Cells were exposed to the effluent of the direct channel with different $O_2$ concentrations added to the He flow fed through the side channel. The concentration of $O_2$ in the He flow through the side channel varied from 0 to 20%. Plasma conditions: $U_{RMS}$ = 230 V, He flow 1.4 slm, $O_2$ concentration 0.6%.

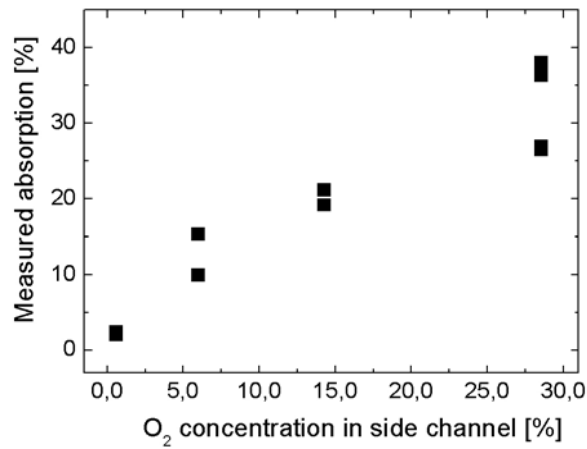

Figure 4: The change of the overall emission intensity in the 115 – 450 nm region, measured by solar blind UV and VUV detector, as function of the $O_2$ concentration in the side channel.

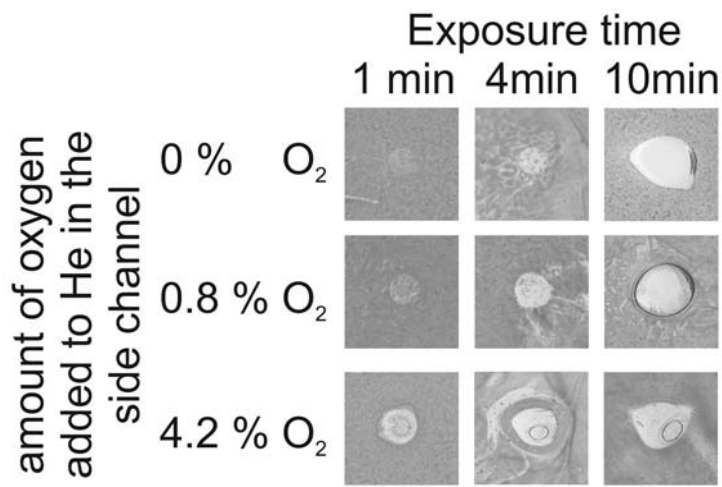

Figure 5: 15 by 15 mm details of photographs of Petri dishes with inhibition zones in an *E. coli* monolayers after 1, 4 and 10 min of treatment. Cells were exposed to the plasma effluent passing through the direct channel allowing for photochemistry events by streaming He with $O_2$ admixture though the side channel. The concentration of $O_2$ in the He flow through the side channel (total flow 2 slm) varied from 0 to 4.2%. Plasma conditions: $U_{RMS}$ = 200 V, He flow 1.4 slm, no $O_2$ added.

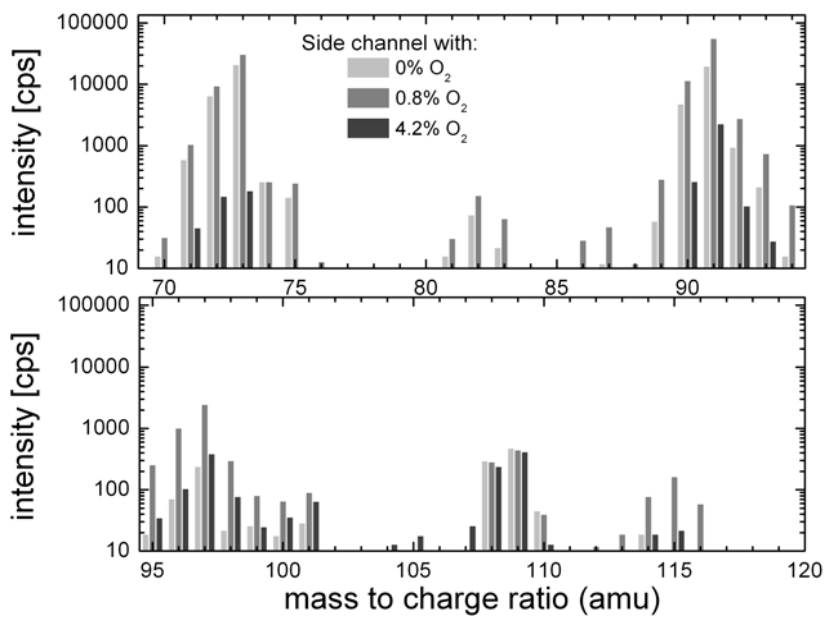

Figure 6: Mass spectrometry measurement of positive ions at 4 mm distance from the end of the direct channel (cf. Figure 1b)) in the mass range 70 to 120 amu. The same conditions as in Figure 5 with three different concentrations of $O_2$ in the He flow through the side channel are compared here

**The table of contents entry:**

**The role of VUV radiation in the inactivation of bacteria with an atmospheric pressure plasma jet**

S. Schneider, J.-W. Lackmann, D. Ellerweg, B. Denis, F. Narberhaus, J. E. Bandow, J. Benedikt

**The role of VUV photons in the atmospheric pressure plasma treatment of bacteria is investigated** by means of a modified version of a micro scale He/O$_2$ atmospheric pressure plasma jet (µ-APPJ) source, so-called X-Jet. The X-Jet allows effective separation of heavy reactive particles such as O or O$_3$ from the plasma-generated photons. The results show that the impact of photochemistry products on bacteria is more effective than direct inactivation of cells by VUV radiation damage and indicate that the combination of water impurity and presence of molecular oxygen at the surface are important for the process.

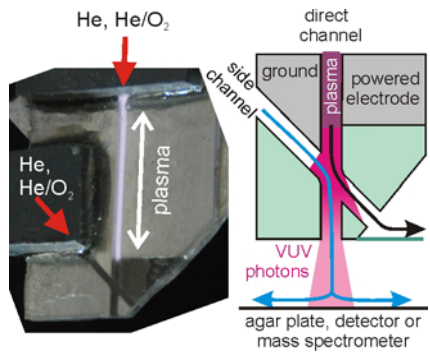